\documentclass[iop]{emulateapj}

\usepackage{color}
\usepackage{amsmath}
\usepackage{natbib}
\newcommand{\xmm}{{\it XMM-Newton}}
\newcommand{\swift}{{\it Swift}}

\shortauthors{C.-P.~Hu et al.}

\begin{document}
\title{Discovery of an X-ray Emitting Contact Binary System 2MASS J11201034$-$2201340}
\author{Chin-Ping Hu\altaffilmark{1,2}, Ting-Chang Yang\altaffilmark{2}, Yi Chou\altaffilmark{2}, L. Liu\altaffilmark{3}, S.-B. Qian\altaffilmark{3}, C. Y. Hui\altaffilmark{4}, Albert K. H. Kong\altaffilmark{5}, L. C. C. Lin\altaffilmark{6}, P. H. T. Tam\altaffilmark{7}, K. L. Li\altaffilmark{8},Chow-Choong Ngeow\altaffilmark{2}, W. P. Chen\altaffilmark{2}, and Wing-Huen Ip\altaffilmark{2}}
\altaffiltext{1}{Department of Physics, University of Hong Kong, Pokfulam Road, Hong Kong; cphu@hku.hk}
\altaffiltext{2}{Graduate Institute of Astronomy, National Central University, Jhongli 32001, Taiwan; yichou@astro.ncu.edu.tw}
\altaffiltext{3}{Yunnan Observatories, Chinese Academy of Sciences, P.O. Box 110, 650011 Kunming, China}
\altaffiltext{4}{Department of Astronomy and Space Science, Chungnam National University, Daejeon, Korea}
\altaffiltext{5}{Institute of Astronomy and Department of Physics, National Tsing Hua University, Hsinchu, Taiwan}
\altaffiltext{6}{Institute of Astronomy and Astrophysics, Academia Sinica, Taiwan}
\altaffiltext{7}{Institute of Astronomy and Space Science, Sun Yat-Sen University, Guangzhou 510275, China}
\altaffiltext{8}{Department of Physics and Astronomy, Michigan State University, East Lansing, MI 48824-2320, USA}


\begin{abstract}
We report the detection of orbital modulation, a model solution, and X-ray properties of a newly discovered contact binary, 2MASS J11201034$-$2201340.  We serendipitously found this X-ray point source outside the error ellipse when searching for possible X-ray counterparts of $\gamma$-ray millisecond pulsars among the unidentified objects detected by the {\it Fermi Gamma-ray Space Telescope}.  The optical counterpart of the X-ray source (unrelated to the $\gamma$-ray source) was then identified using archival databases.   The long-term CRTS survey detected a precise signal with a period of $P=0.28876208(56)$ days.  A follow-up observation made by the SLT telescope of Lulin Observatory revealed the binary nature of the object. Utilizing archived photometric data of multi-band surveys, we construct the spectral energy distribution, which is well fitted by a K2V spectral template.   The fitting result of the orbital profile using the Wilson--Devinney code suggests that 2MASS J11201034-2201340 is a short-period A-type contact binary and the more massive component has a cool spot.  The X-ray emission was first noted in observations made by \swift\, then further confirmed and characterized by an \xmm\ observation. The X-ray spectrum can be described by a power law or thermal Bremsstrahlung.  Unfortunately, we could not observe significant X-ray orbital modulation. Finally, according to the spectral energy distribution, this system is estimated to be 690 pc from Earth with a calculated X-ray intensity of $(0.7-1.5)\times 10^{30}$ erg s$^{-1}$, which is in the expected range of an X-ray emitting contact binary.  
\end{abstract}


\section{Introduction}
A W UMa-type system is a contact binary system where both components share a common envelope and are typical main sequence stars with similar surface temperatures.  Astronomers have been aware of the optical variability of prototypical W UMa-type systems was noticed for more than a century \citep{Muller1903}.  To date,  there are thousands of known W UMa-type variables were found.  The spectral type of this kind of system usually ranges from A to K, with a period that ranges from 0.2 to 1.4 days and variability amplitudes typically less than 1 magnitude.  Contact binaries can be further classified into two major types: A and W \citep{Binnendijk1970}. The primary minimum in the folded light curve of an A-type contact binary is caused by the less massive component transiting the more massive one; otherwise, the contact binary is a W-type.   In general, an A-type system has a relatively longer orbital period ($P \gtrsim 0.3$ day), a lower mass ratio ($q\lesssim0.3$), and an earlier spectral type (typically from A to G).  On the other hand, a W-type system usually has a later spectral type.   A- and W-type contact binaries are possibly related with respect to their evolution.  For example, \citet{Hilditch1988} suggested that a W-type system evolves to an A-type system; however, \citet{Gazeas2006} proposed an opposite evolutionary track.   


W UMa-type systems are expected to have high chromospheric activity and coronal emission; hence, some systems are strong X-ray emitters, e.g., VW Cep \citep{Carroll1980, Huenemoerder2006}.  The strength of the X-ray emission is related to the binary orbital period and the spectral type \citep{Stepien2001, Chen2006}.  Chromospheric activity is related to the presence of cool spots, which cause asymmetry, known as the O'Connell effect \citep{OConnell1951,Wilsey2009}, in the orbital profile.  In addition, the variability of the H$\alpha$ equivalent width along with the orbital phase is an indication of chromospheric activity \citep{Kaszas1998}.  A detailed investigation of the X-ray timing and spectral variability, as occurred with the brightest contact binary VW Cep \citep{Huenemoerder2006}, can reveal the position and geometry of the corona. 

\begin{figure*}
\plotone{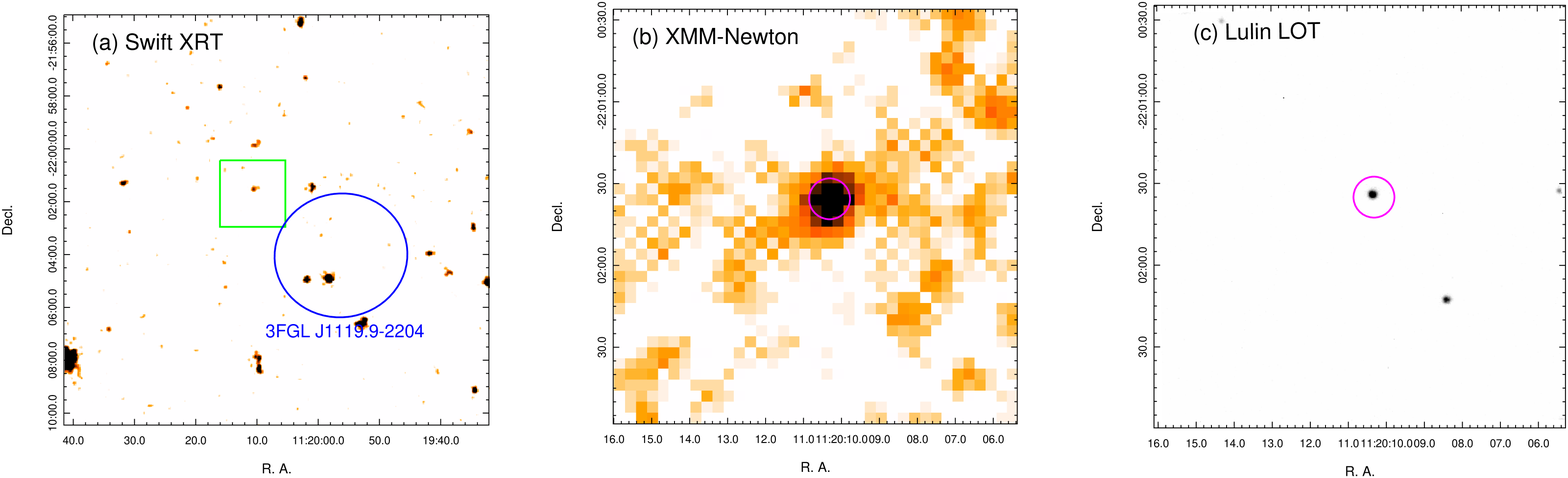} \caption{(a) \swift\ image of the field containing 2MASS J11201034$-$2201340 and 3FGL J1119.9$-$2204.  This image is smoothed with a Gaussian kernel of $\sigma=3"$ to enhance the visibility of faint sources.   The blue circle encompasses the 95\% error ellipse of 3FGL J1119.9$-$2204 and the green box is the region of the cropped \xmm\ and  Lulin LOT r'-band images in the center and right panels, respectively. (b) Enlarged view of the \xmm\ image.  The center of the cyan circle is located at the \xmm\-determined position, with coordinates R.A. = 11$^h$20$^m$10.32$^s$ and Decl. = $-$22$^{\circ}$01'35.6" and the diameter of the circle is the \xmm\ half-energy width. (c) The r'-band image of the same field in the center panel taken by the LOT. \label{xmm_lot}}
\end{figure*}

We present the detection of the contact binary 2MASS J11201034$-$2201340 in X-ray, optical, and infrared bands in Section \ref{detection}.  Section \ref{optical_obs} presents the results of optical band analysis, including the determination of the orbital period and the spectral energy distribution.  We also present the photometric solution to determine the basic physical parameters of this system, utilizing orbital profile fitting.  Section \ref{xray_obs} presents X-ray timing and spectral behaviors determined from the results of X-ray data analysis, e.g., the non-detection of orbital modulation and the non-thermal spectral nature. We also estimate X-ray intensity and discuss the relationship between X-ray and optical luminosities. Finally, we summarize our results and future aspects in Section \ref{summary}.


\section{Source Detection}\label{detection}

\subsection{\swift\ Observations}
After the detection of 1FGLJ1119.9$-$2205 (also named 2FGL J1120.0$-$2204 and 3FGL J1119.9$-$2204), the \swift\ XRT took 27 exposures of this field for a total exposure time of $\sim 67$ ks between 2010 and 2013.  \citet{Hui2015} studied all the \swift\ observational data, found two millisecond pulsar candidates within the 95\% error ellipse of 3FGL J1119.9$-$2204.  On the other hand, several uncataloged X-ray point sources outside the error ellipse also were detected but their properties were not further investigated.  2MASS J11201034$-$2201340, which has relatively faint X-ray emission, is one of the outliers unrelated to the $\gamma$-ray source 3FGL J1119.9$-$2204.

The energy range of XRT is 0.2--10 keV, the pixel scale is 2.36\arcsec, and the full-width at half maximum (FWHM) of the point spread function (PSF) is roughly 7\arcsec\ in 1.5 keV (or a half-power diameter of 18\arcsec\!).  All the data, including two target IDs (41371 and 49351), were used to determine the X-ray positions of the point sources.  We extracted photon events and X-ray images from the standard products of all the XRT observations using {\it xselect} version 2.4.  The point sources were detected using the {\it detect} task of the multi-mission X-ray image analysis program {\it XIMAGE}, for which the signal-to-noise threshold was set to 3.0. The position and corresponding uncertainty were determined using the {\it xrtcentroid} task.  We found an X-ray point source located at R.A. = 11$^h$20$^m$10.65$^s$ and Decl. = $-$22\arcdeg 01\arcmin 30.7\arcsec\ with a 90\% uncertainty of $\sim$ 7.6\arcsec. Figure \ref{xmm_lot}(a) is the \swift\ image of the field containing 2MASS J11201034$-$2201340 and 3FGL J1119.9$-$2204.

\subsection{\xmm\ Observation}
2FGL J1120.0$-$2204 was observed by the \xmm\ observatory on 2014 June 14 for a total exposure time of $\sim 70$ ks (\dataset[ADS/Sa.XMM#obs/0742930101]{ObsID 0742930101}).  All three detectors of the European Photon Imaging Camera (EPIC) were used in this observation.  The MOS1 and MOS2 detectors were operated in full-frame mode with a timing resolution of 2.6 s and an on-axis PSF FWHM of 6\arcsec\ (half-energy width 13.6"). The pn detector was operated in timing mode with an extreme timing resolution of 0.03 ms. In this mode, all X-ray photons are compressed in one dimension.  However, this observation was performed to investigate the timing properties of the millisecond pulsar candidate 2FGL J1120.0-2204. The pn data were useless because the target is far from the aim point of the observation.  The optical/UV monitor also was used in this observation but our target was outside the field of view and the optical/UV monitor data were unavailable.  We applied the pipeline task {\it emproc} of the XMM Science Analysis Software (XMMSAS version 15.0.0)  program to the MOS data using the latest instrumental calibration database.  Events with patterns $<12$ were adopted in this research, and the flaring background was filtered out when the entire count rate was $>3.5$ counts s$^{-1}$.  We performed source detection using the maximum likelihood fitting with the aid of the XMMSAS task {\it edetect\_chain}  and the signal-to-noise threshold set to $4\sigma$.

We found a source located at R.A. = 11$^h$20$^m$10.32$^s$ and Decl. = $-$22\arcdeg 01\arcmin 35.6\arcsec\ with a 90\% uncertainty of 2\arcsec.  The detection likelihood is 34.5, which corresponds to a 8 $\sigma$ significance level.  This source is likely a point source because the likelihood of source extent is not determined. Figure \ref{xmm_lot}(b) shows the \xmm\ image of the region containing 2MASS J11201034$-$2201340 where a point source is clearly seen.  

\begin{figure*}
\plottwo{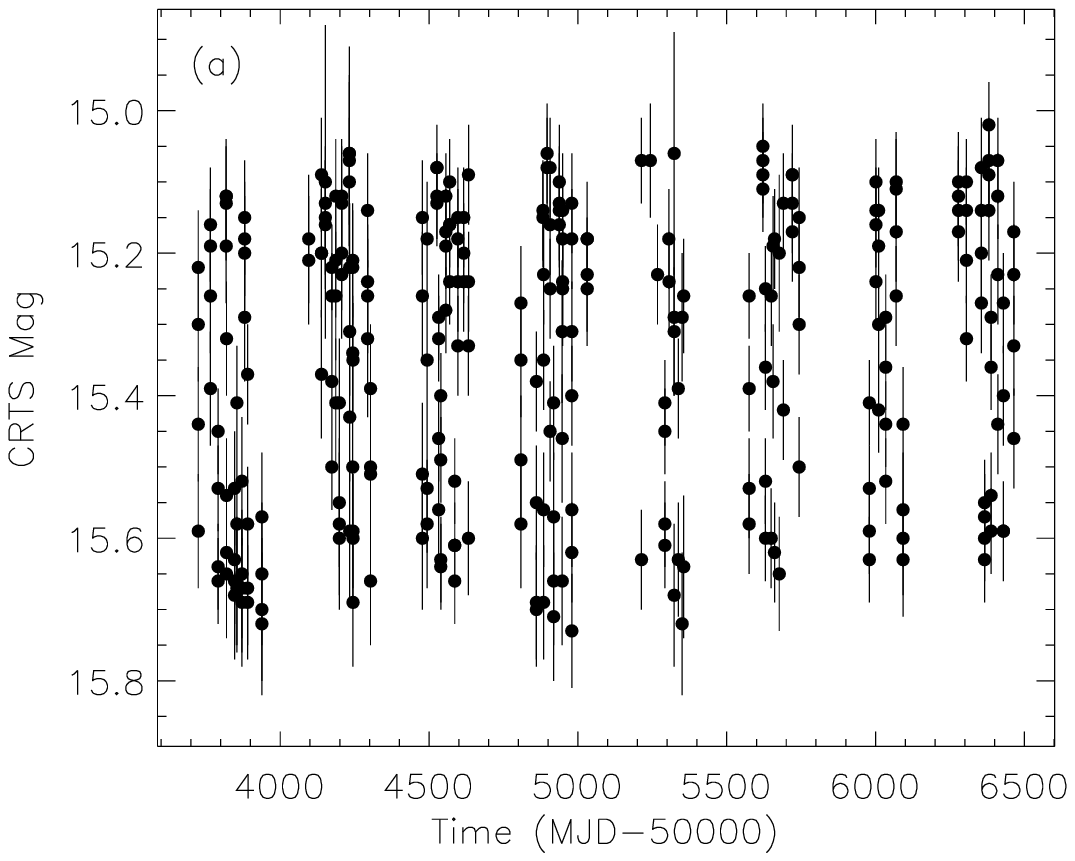}{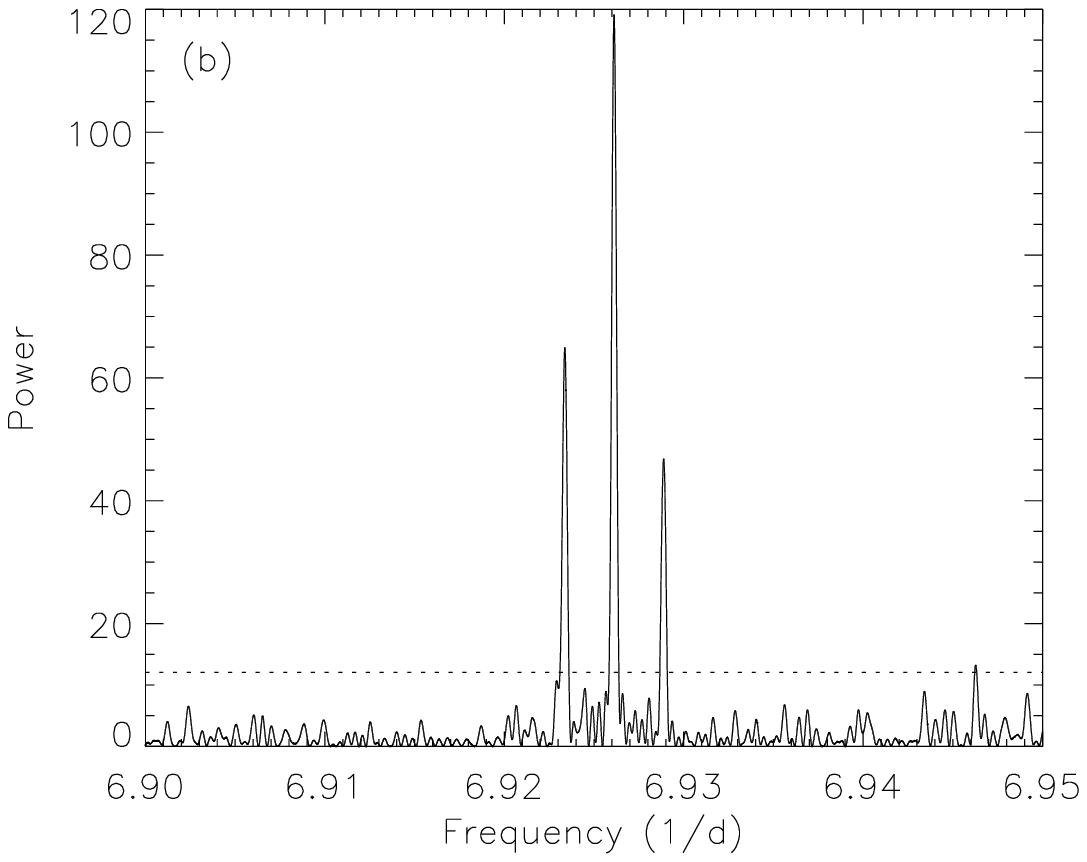} \caption{The CRTS light curve (a) and the corresponding Lomb--Scargle power spectrum (b) of 2MASS J11201034$-$2201340.  A significant peak located at $f=6.926$ 1/day and its two associated 1-year aliases are clearly seen. The 3-$\sigma$ white noise level is indicated by the dashed line.  \label{lsp}}
\end{figure*}

\subsection{Optical and Infrared Counterparts}
The field around the X-ray source was surveyed in optical and infrared bands by  USNO-B-1.0, the Two Micron All Sky Survey (2MASS), and the Wide-field Infrared Survey Explorer (WISE).  All three catalogs are well organized by the the NASA/IPAC Infrared Science Archive (IRSA). Each catalog has only one source that is within the PSF of the \xmm: USNO-B1.0 0679$-$0311979 at R.A. = 11$^h$20$^m$10.36$^s$ and Decl. = $-$22\arcdeg 01\arcmin 34.14\arcsec, 2MASS J11201034$-$2201340 at R.A. = 11$^h$20$^m$10.35$^s$ and Decl. = $-$22\arcdeg 01\arcmin 34.04\arcsec, and WISE J112010.35$-$220134.1 at R.A. = 11$^h$20$^m$10.35$^s$ and Decl. = $-$22\arcdeg 01\arcmin 34.16\arcsec.  2MASS J11201034$-$2201340 is 5.3\arcsec\ from the \swift\ determined X-ray source, and 1.5\arcsec\ from the position determined by \xmm. Hence, 2MASS J11201034$-$2201340 is likely to be the optical counterpart of the X-ray source because the angular separation between the optical and X-ray positions well reside in the uncertainties of \swift\ and \xmm. 

We also observed the field with the Lulin One-Meter Telescope (LOT), located at Central Taiwan, using the ALTA U42 CCD with SDSS r'-band and g'-band filters.  This configuration has a pixel scale of 0.32\arcsec\ and a limiting magnitude of $\sim$19.5 at the detection level of 5 $\sigma$.  Figure \ref{xmm_lot}(c) shows the cropped r'-band image.  Only one source is detected in the region of the point spread function of \xmm.

\section{Detailed Optical Investigation}\label{optical_obs}

\begin{figure}
\plotone{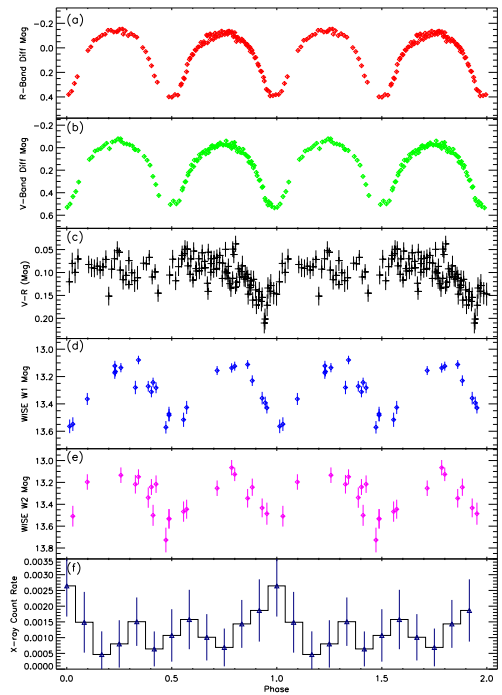} \caption{The multi-band light and color curves of 2MASS JJ11201034$-$2201340 folded according to a period of 0.288 day. (a) and (b) Folded R-band and V-band light curves obtained by SLT, (c) V $-$ R color variability, (d) and (e) folded W1-band and W2-band light curves obtained by WISE, and (f) folded X-ray light curve obtained by \xmm\ with energy of 0.2 -- 10 keV. \label{fold_lc}}
\end{figure}

\subsection{Timing Analysis}
To investigate the temporal variation of 2MASS J11201034$-$2201340, we first collected the multi-epoch photometric data obtained by the Catalina Real-Time Transient Survey (CRTS) to search for periodicity. The CRTS data set contains the observations made by the Siding Springs Survey (SSS) 0.5-m Schmidt telescope (218 exposures) and by the Catalina Sky Survey (CSS) 0.7-m Schmidt telescope (76 exposures).  We found a significant periodicity with a period of 0.144 day by applying the Lomb-Scargle periodogram \citep{Scargle1982, Horne1986}.  Figure \ref{lsp} shows the light curve from the CRTS and the power spectrum.  One-year aliases due to the observational window function also were detected in the power spectrum.

To check the robustness of the detected result and the possibility that the true periodicity is twice the detected value, we observed the field using the R and V bands of the 0.4-m Super Light Telescope (SLT) at the Lulin Observatory on 2015 April 24--27.  The SLT was equipped with an Andor iKon-L 936 CCD with a field of view of $28 \times 28$ \arcmin\ and a pixel scale of 0.82\arcsec.  The exposure time for each frame was 5 min, and the observation lasted $\sim 5$h on April 24, 27, and 2.2 h on April 26.  The R band had 117 exposures and the V band had 115.  For  comparing the phase alignment of the CRTS data, the SLT data, and the data from other wavelengths, the observation times were corrected from the Earth to the solar barycenter.  The CCD reduction package of the {\it Image Reduction and Analysis Facility} ({\it IRAF}) was used for standard image reduction, including bias, dark current, and flat-field corrections.  We then compared the instrumental magnitude of 2MASS J11201034$-$2201340 with that of four comparison stars in the same field to obtain differential photometric data. The light curves clearly showed variability on a time scale of several hours. After folding both the light curves and the B$-$V color curve using 0.144 and 0.288 day, respectively, we found that the color seemed to vary with a period of 0.288 day (see Figure \ref{fold_lc}).  The true period of $0.28876208\pm5.6\times10^{-7}$ day is twice that detected by the Lomb-Scargle periodogram from the CRTS light curve.   The uncertainty was estimated by a combination of the limited time span and photon statistical error.  The uncertainty caused by limited time span and the strength of the signal was estimated using Equation 3 in \citet{LevineBC2011} and the statistical error was estimated using a $10^4$ times Monte Carlo simulation.

From the archived multiepoch photometry data of WISE observations of this field, we found 52 W1-band (3.4$\mu$m) and 42 W2-band (4.6 $\mu$m) significant detections and folded them with the best-determined periodicity.  Figure \ref{fold_lc}  shows that there is still modulation and it is coherent in the mid-infrared bands. 

\subsection{Spectral Energy Distribution}
We further investigated the physical properties of this system using currently available broad-band spectral energy distribution (SED) data.  For the optical band, we used the photometric B, R, and I band data in the USNO-B 1.0 catalog, where $B1=16.4$, $B2=16.03$, $R1=14.36$, $R2=14.96$, and $I=14.55$.  To describe the spectral behavior of a contact binary system, it is better to use one of the flux minimum in the folded light curve that has all the photons from the primary component.  However, the amount of current data is insufficient because only one or two measurements were made for individual bands.   Therefore, we took the average of the detected magnitudes for the B and R bands for which there were two measurements, and applied a typical photometric uncertainty of 0.3 mag \citep{Monet2003}. The near-infrared data were from 2MASS in the J, H, and Ks bands, where $J=13.875\pm0.03$, $H=13.354\pm0.03$, and $Ks=13.276\pm0.04$.  The mid-infrared data were from WISE in the W1 and W2 bands, where  $W1=13.286\pm0.025$ and $W2=13.331\pm0.031$.  

\begin{figure}
\plotone{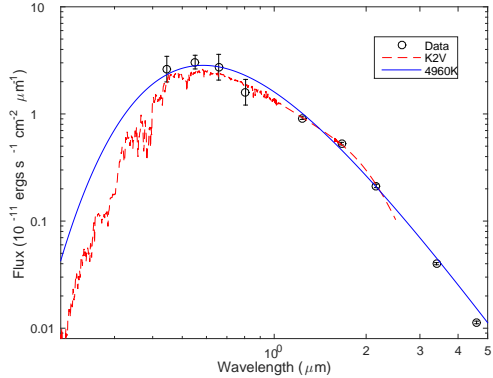} \caption{Spectral energy distribution of 2MASS J11201034$-$2201340. The data points were obtained from the USNO-B 1.0 catalog, the 2MASS catalog, and the WISE All-Sky Point Source catalog. The red dashed line is the best-fit K2V spectral template and the blue line is the corresponding blackbody radiation spectrum. \label{sed}}
\end{figure}

The SLT did not observe standard stars so calibration in the V and R bands was not possible.  Instead, to estimate the apparent magnitude of 2MASS J11201034$-$2201340, we used the cataloged star TYC 6090-207-1, which is a bright star with a coordinate of R.A. = 11$^h$19$^m$56.983$^s$ and Decl. = $-$22\arcdeg04\arcmin43.94\arcsec and magnitudes of $B=13.09\pm0.26$ and a $V=12.20\pm0.16$ \citep{Hog2000}.  By scaling of the brightness of 2MASS J11201034$-$2201340 and TYC 6090-207-1, we estimated the V-band magnitude of 2MASS J11201034$-$2201340 as $15.37\pm0.16$. 

Before constructing the SED, the measured magnitudes underwent extinction correction using the online Galactic dust extinction tool provided by IRSA.  We assumed that the relationship between the extinction ($A_V$) and reddening ($E(B-V)$) is $A_V=3.1E(B-V)$ \citep{Guver2009}. We estimated the extinction for individual bands according to the latest measurements provided by \citet{SchlaflyF2011}. Figure \ref{sed} shows the constructed SED of 2MASS J11201034$-$2201340. We compared the SED with various stellar spectral templates presented in \citet{Pickles1998}  and found that our SED fit that of a K2V star that has a typical surface temperature of 4960 K. This indicates that the major component of 2MASS J11201034$-$2201340 is a late-type star.    

\begin{figure}
\plotone{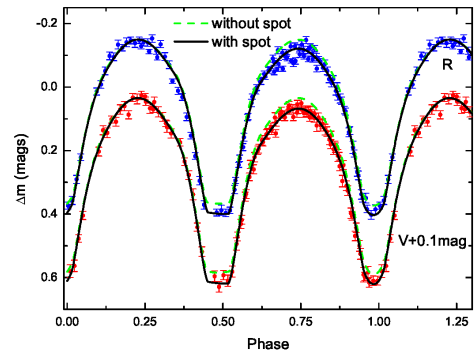} \caption{The observed folded R-band light curve (blue), folded V-band light curve (red) with a shift of 0.1 mag, and the corresponding best-fit contact binary model estimated using the Wilson--Devinney code.  The green dashed line is the best-fit model without a cool spot and the black line is the best-fit model with a cool spot. \label{profile_fitting}}
\end{figure}

\begin{table*}
\centering
\begin{small}
\caption{Photometric solutions for 2MASS J11201034$-$2201340.}
\label{photometric_table}
\begin{tabular}{lclcl}
\hline
Parameters              &  Photometric elements  &  Errors     &  Photometric elements  &  Errors      \\
                                                                                                       \\
                        &  without cool spot  &             &  cool spot solutions  &              \\
\hline \hline                                                                                          
$g_1=g_2$               &     0.32               & assumed     &     0.32               & assumed      \\
$A_1=A_2$               &     0.50               & assumed     &     0.50               & assumed      \\
$x_{1bol},x_{2bol}$   &     0.299,0.299        & assumed     &     0.299,0.299        & assumed      \\
$y_{1bol},y_{2bol}$   &     0.396,0.396        & assumed     &     0.396,0.396        & assumed      \\
$x_{1V},x_{2V}$         &     0.581,0.581        & assumed     &     0.581,0.581        & assumed      \\
$y_{1V},y_{2V}$         &     0.249,0.249        & assumed     &     0.249,0.249        & assumed      \\
$x_{1R},x_{2R}$         &     0.356,0.356        & assumed     &     0.356,0.356        & assumed      \\
$y_{1R},y_{2R}$         &     0.413,0.413        & assumed     &     0.413,0.413        & assumed      \\
$T_1$                   &     4960K              & fixed       &     4960K              & fixed        \\
Phase Shift             &     $-0.0153$          & $\pm0.0005$ &     $-0.0155$           & $\pm0.0004$  \\
$q$                     &     0.291              & $\pm0.006$  &     0.317               & $\pm0.004$  \\
$\Omega_1=\Omega_2$     &     2.4253             & $\pm0.0126$ &     2.4735              & $\pm0.0111$ \\
$\Omega_{in}$           &     2.4464             & --          &     2.5026              & --          \\
$\Omega_{out}$          &     2.2645             & --          &     2.3060              & --          \\
$T_2$                   &     5125K              & $\pm20$K    &     5124K              & $\pm16$K     \\
$i$                     &     $89^{\circ}.4$     & $\pm3.3$    &     $88^{\circ}.5$     & $\pm2.7$     \\
$L_1/(L_1+L_2)(V)$      &     0.7180             & $\pm0.0051$ &     0.7019             & $\pm0.0046$  \\
$L_1/(L_1+L_2)(R)$      &     0.7259             & $\pm0.0045$ &     0.7100             & $\pm0.0042$  \\
$\theta$ ($^\circ$)     &                        &             &     86.5               & $\pm2.6$     \\
$\psi$ ($^\circ$)       &                        &             &     92.2               & $\pm4.7$     \\
$\Omega(^\circ)$      &                        &             &     11.9               & $\pm1.2 $    \\
$T_s/T_*$               &                        &             &     0.80               & $\pm0.09$    \\
$f$                     &     $11.6\,\%$         & $\pm6.9\,\%$&     $14.8\,\%$         & $\pm5.7\,\%$ \\
\hline
\end{tabular}
\end{small}
\end{table*}

\subsection{Photometric Solution}\label{Timing_Solution}
To determine the physical nature of this contact binary system using the available data, we fitted the observed V-band and R-band light curves using the Wilson--Devinney code \citep{Wilson1971, Wilson1979, Wilson1990, Wilson1994} to obtain a photometric solution.  The surface temperature of the more massive star ($T_1$) was fixed at 4960 K, as the SED fitting suggested.  The gravity darkening parameters were fixed at $g_1=g_2=0.32$, according to the prediction made by \citet{Lucy1967}, and the bolometric albedos were fixed at $A_1=A_2=0.5$ \citep{Rucinski1969} on the basis of the assumption that both components have convective envelopes. The limb-darkening coefficients for bolometric and individual bands ($x_{1bol}$, $x_{2bol}$, $y_{1bol}$, $y_{2bol}$, $x_{1V}$, $x_{2V}$, $y_{1V}$, $y_{2V}$, $x_{1R}$, $x_{2R}$, $y_{1R}$, and $y_{2R}$) were estimated using the values provided by \citet{vanhamme1993} (see Table \ref{photometric_table}).  The remaining free parameters were the mass ratio ($q$), the inclination angle of the orbital plane ($i$), the surface temperature of the less massive component ($T_2$), the phase shift of V and R bands, the dimensionless potential of the more massive star ($\Omega_1=\Omega_2$), and the relative luminosity ($L_1$ and $L_2$) of each star in the V and R bands.   

Because the spectral mass ratio could not be determined using the radial velocity method, we searched for a proper initial guessing value before fitting the light curves.  We estimated the sum of the weighted square deviations ($\Sigma$) with respect to  different trial values of $q$, ranging between 0.1 and 6.  A minimum of $\Sigma$ was clearly seen at $q=0.3$, so we chose that as the initial $q$ values for the fitting. Moreover, the profile seemed to be asymmetric, so we introduced a cool (dark) spot in the fitting to compare against the model without a cool spot. The cool spot model included three additional parameters: the colatitude ($\theta$), longitude ($\psi$), and radius ($\Omega$) of the spot. The best-fit parameters are given in Table \ref{photometric_table} and the model light curves are shown in Figure \ref{profile_fitting}. 

The magnitude of the secondary minimum (phase $\sim$0.5), which is not much different from that of the primary in the folded light curve, indicates that the less massive star is completely obscured and only the more massive star contributes to the flux.  The V-band magnitude of the secondary minimum is estimated as $15.75\pm0.16$ in this measurement. Therefore, the distance can be roughly estimated as $690 \pm 50$ pc after considering the extinction and assuming that the more massive component is a K2V star. 

The result of the fitting indicates that 2MASS J11201034$-$2201340 is a short-period A-type contact binary with a nearly edge-on orbital plane.  It is necessary that there be a cool spot on the more massive component to explain the O'Connell effect.  In this scenario, the best-fit mass ratio is $q=0.317$ and a fill-out factor is estimated to be $f=14.8 \%$, which suggests a low degree of over contact.  From the averaged value of the phase shift, we proposed a linear ephemeris for the primary minimum defined as
\begin{equation*}
T_{{\rm primary}}={\rm MJD (TDB) } 53791.3984(1) + 0.28876208(56)\times N,
\end{equation*}
where $N$ is the number of cycle count.

\section{Results from X-ray Observations}\label{xray_obs}

\begin{figure*}
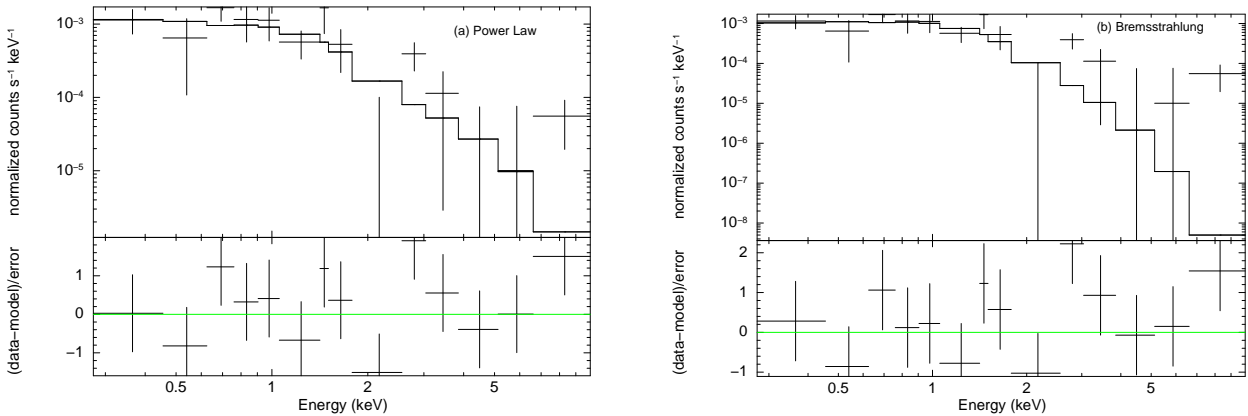

\centering
\columnwidth=0.7\columnwidth
\includegraphics*[width=0.9\columnwidth,angle=270]{powerlaw1.ps}
\hfil
\includegraphics*[width=0.9\columnwidth,angle=270]{bremss1.ps}
\caption{(a) The X-ray spectral data (crosses) and the best-fit power-law model (steps) of 2MASS J11201034$-$2201340 obtained with \xmm. The lower panel shows the deviation of each data point from the best-fit model. (b) The same as (a) for the thermal Bremsstrahlung model.  \label{powerlaw_spectrum}}
\end{figure*}

\subsection{X-ray Profile and Spectrum}
Investigation of the timing and spectral properties of the X-ray emission from 2MASS J11201034$-$2201340 detected by \xmm\ will add to our understanding of the nature of this system.  After filtering out the flaring background, the effective exposure time was $\sim 57$ ks.  We extracted 238 X-ray photons from a 20-arcsec-radius circular region centered at the detected centroid of 2MASS J11201034$-$2201340 (see Figure \ref{xmm_lot}).   Because the effective exposure time encompassed only approximately two orbital cycles of 2MASS J11201034$-$2201340, we needed to carefully examine background variability.  We selected ten backgrounds with the same area in the source-free region on the same chip and found that the background counts were $\sim 150$.  Therefore, only $\sim 90$ X-ray photons that originated from the source were available for use in a marginal investigation of the timing and spectral analysis.   First, we folded individually the source and all the backgrounds according to the best period determined in the optical band.  Then, we obtained a mean folded light curve by averaging all the folded background light curves to obtain the fluctuation of the background in the phase domain.  The clean folded light curve shown in Figure \ref{fold_lc} (f) was obtained by subtracting the background profile from the folded source light curve.  Another way compute the background contribution is by weighting a larger source-free region, e.g., a circle with an 80-arcsec radius or a large annulus around the source. Both methods yield similar results.  The folded light curve showed no detection of significant X-ray variability.  We used different bin sizes and applied the $\chi^2$ test to the folded profile and found that the detection significance in all cases was $<1\sigma$.  The lack of detection of variability means that the geometry of the X-ray emission area may differ from that of the optical area.  

The high-energy emission mechanism can be investigated by X-ray spectral analysis, despite the small number of X-ray photons. The source selection criterion was the same as that for the X-ray timing analysis above, while we used an 80-arcsec-radius circle around a source-free region to estimate the background spectrum.  The response matrix was created using the XMMSAS task {\it rmfgen} and the ancillary response file was created using {\it arfgen}.  The X-ray photons were further grouped to have at least 15 counts per spectral bin.  No pile-up issue was addressed for this faint source. 

The X-ray spectral fitting was achieved using {\tt XSPEC v12.9.0}. The X-ray emission from a contact binary system may come from a hot coronal plasma and may be represented with a thermal or non-thermal model. We tried two typical spectral models: non-thermal power law ({\tt XSPEC} model {\tt powerlaw}), and the thermal Bremsstrahlung ({\tt XSPEC} model {\tt bremss}) to fit the X-ray spectrum.  The spectral fitting was calculated within the energy range 0.2--10 keV.  We first set the $N_{\rm{H}}$ to be a free parameter and yielded a best-fit value of $2.7 \times 10^{-20}$ cm$^{-2}$, which is not very far from $3.65 \times 10^{-20}$ cm$^{-2}$ derived from Leiden/Argentine/Bonn (LAB) Survey \citep{Kalberla2005}. However, the uncertainty is $5 \times 10^{-21}$ cm$^{-2}$, which is too large to well constrain the $N_{\rm{H}}$ value.  Therefore, we fixed the $N_{\rm{H}}$ value at the galactic one of $3.65 \times 10^{-20}$ cm$^{-2}$ in the following analysis.  The power-law model yielded an acceptable fitting with a photon index of $2.4\pm0.5$ and a $\chi_{\nu}^2=1.09$.  The uncertainties of spectral parameters are estimated within the 90 \% confidence interval.  On the other hand, the thermal Bremsstrahlung model result in an equally good fitting with a slightly larger $\chi_{\nu}^2$ of 1.15 and a plasma temperature of $kT=0.8^{+1.4}_{-0.3}$\,keV.  Both of the single component model can describe the X-ray spectrum of 2MASS J11201034$-$2201340 very well and no additional components are required.  The best-fit parameters for the two models are given in Table \ref{spectral_fitting}, and the spectral data and corresponding best-fit models are shown in Figure \ref{powerlaw_spectrum}. The X-ray flux in the 0.3--10 keV energy range was estimated as $1.7^{+0.9}_{-0.5}\times10^{-14}$ erg cm$^{-2}$ s$^{-1}$. 

\begin{table}
\centering
\caption{The best-fit parameters of two different models for the X-ray spectrum of 2MASS J11201034$-$2201340 obtained with \xmm. }
\label{spectral_fitting}
\begin{tabular}{ccc}
\hline 
 & Power Law & Bremsstrahlung\tabularnewline
\hline 
\hline 
$\Gamma$ & $2.4\pm0.5$ & \nodata \\
$kT$ (eV) & \nodata & $0.8^{+1.4}_{-0.3}$ \\
Normalization & $4.2_{-1.3}^{+1.2}\times10^{-6}$ & $1.1_{-0.6}^{+1.0}\times 10^{-6}$ \\
$\chi_{\nu}^2$ (dof) & 1.09(12) & 1.15 (12)\\
\hline 
\end{tabular}
\end{table}

\subsection{Indications from X-ray Emission}
Because the fitting of the optical folded light curve suggested the presence of a cool spot, the existence of X-ray emission suggests chromospheric activity.  The result of the X-ray spectral fitting also suggests a possible non-thermal or thermal Bremsstrahlung origin for the X-ray emission.   However, the X-ray orbital profile shows no significant variations, but this may be due to an insufficient number of X-ray photons. Considering the stability of star spots, it is possible that their position and size in these two observations are quite different.  To investigate the origin of the X-ray emission and its connection to the star spot, simultaneous X-ray and optical observations using an X-ray telescope with a large effective area are necessary.

Although only $\sim$ 30X-ray photons were detected in all the \swift\ observations, we were able to estimate a crude X-ray spectrum and the corresponding X-ray flux by using the online tool provided by UK \swift\ Science Data Centre. The flux of X rays between 0.3 and 10 keV determined by \swift\ was $1.0^{+0.6}_{-0.5}\times10^{-14}$ erg cm$^{-2}$ s$^{-1}$, which is consistent with the value obtained by a single \xmm\ observation ($1.7^{+0.9}_{-0.5}\times10^{-14}$ erg cm$^{-2}$ s$^{-1}$). This may indicate that no significant X-ray flare was detected during the \swift\ observations.  

\begin{figure}
\plotone{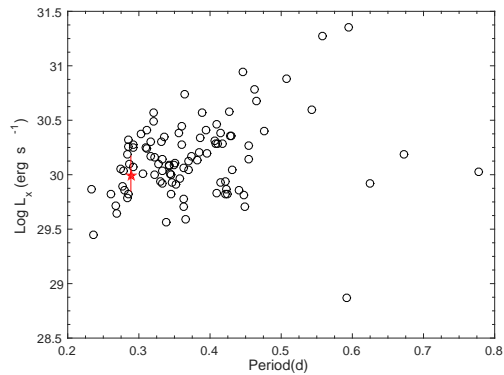} \caption{The relation between the X-ray luminosity and orbital period for contact binaries.  The red star denotes 2MASS J11201034$-$2201340 and the open circles are historical data adopted from \citet{Chen2006}. \label{lx_porb}}
\end{figure}

Using the distance estimated from the SED, the X-ray intensity was estimated to be $(0.7-1.5)\times 10^{30}$ erg s$^{-1}$. Factoring in the relationship between the X-ray intensity and the orbital period \citep{Chen2006}, 2MASS J11201034$-$2201340 is within a reasonable range on the X-ray intensity vs.~orbital period plot (see Figure \ref{lx_porb}). \citet{Stepien2001} categorized W UMa-type stars into hot $[(B-V)_0\leq0.6]$ and cool $[(B-V)_0>0.6]$ groups.  The X-ray flux seems to positively correlate with the color index for the hot group, whereas it reaches a constant value after $(B-V)_0>0.6$.  Judging from the SED fitting, 2MASS J11201034$-$2201340 is likely to be in the cool group.  The ratio of the X-ray and bolometric fluxes, $\log(L_x/L_{bol})$, is roughly estimated to be between $-3$ and $-3.3$, which is reasonably close to the value presented in \citet{Stepien2001}.  Accurate multi-band photometric observations and standard star calibration are required to investigate the nature of this system in detail.

\section{Summary}\label{summary}
While searching for the X-ray and optical counterparts of millisecond pulsar candidates, we serendipitously detected the modulation of the uncataloged X-ray-emitting contact binary 2MASS J11201034$-$2201340. We presented a multi-wavelength investigation of the object because it is detectable from infrared to X-ray bands. We first detected 2MASS J11201034$-$2201340 after we combined all the \swift\ observations of  2FGL J1120.0$-$2204.  The exact X-ray position was further confirmed by an \xmm\ observation.  For the optical band, the long-term CRTS light curve and SLT color measurements confirmed that the orbital period of this system is 0.288 day, which is between the values for A-type and W-type contact binaries. Photometric measurement data in the USNO-B 1.0, 2MASS, and WISE catalogs indicate that the broad-band SED of 2MASS J11201034$-$2201340 is that of a K-type star. We used the Wilson--Devinney code to fit the orbital profile in the R- and V-band follow-up light curves observed by the SLT and showed that this system is an A-type contact binary with a mass ratio of $\sim0.3$.  Furthermore, we proposed a cool spot to explain the asymmetric orbital profile.  In addition to the optical periodicity, the X-ray emission hints at the X-ray origin of this object.  Although the number of X-ray photons was insufficient to obtain a significant X-ray orbital profile, the X-ray spectrum is likely to be nonthermal and thus can be linked to chromospheric activity and star spots.  The X-ray flux, orbital period, and color index of 2MASS J11201034$-$2201340 all point to it being a typical contact binary.  Additional multi-band observations with standard star calibrations and long-time-baseline monitoring should conclusively determine the physical properties of this object in detail.

\acknowledgments

This work made use of the data collected by \xmm, an ESA science mission with instruments and contributions directly funded by ESA Member States and the USA (NASA).  The \swift\ data was supplied by the UK \swift\ Science Data Centre at the University of Leicester.  The orbital ephemeris is obtained using the data collected by the CRTS survey, which is supported by the U.S. National Science Foundation under grants AST-0909182. The LOT and SLT are maintained by the Graduate Institute of Astronomy, National Central University.  The near infrared data is collected by the 2MASS, which is a joint project of the University of Massachusetts and the Infrared Processing and Analysis Center/California Institute of Technology, funded by the NASA and the National Science Foundation. The mid-infrared data was provided by the WISE, which is a joint project of the University of California, Los Angeles, and the Jet Propulsion Laboratory/California Institute of Technology, funded by NASA.

C.-P.~H. is supported by the NSC 101-2119-M008-007-MY3 grant from the Ministry of Science and Technology (MOST) of Taiwan and an ECS grant of Hong Kong Government under HKU 709713P.  T.-C.~Y. and W.-H.~I. are supported by the MOST grant NSC 101-2119-M008-007-MY3 and MOST 104-2119-M-008-024.  Y.~C. is supported by the MOST grant NSC 102-2112-M-008-020-MY3.  L.~L. and S.-B.~Q. is supported by the Science Foundation of Yunnan Province (2012HC011) and the Chinese Natural Science Foundation (No.11133007 and No. 11325315).  C.~Y.~H. is supported by the National Research Foundation of Korea through grant 2014R1A1A2058590. A.~K.~H.~K. is supported by the MOST grant 103-2628-M-007-003-MY3. P.~H.~T. is supported by the One Hundred Talents Program of the Sun Yat-Sen University. C.-C.~N. is supported by the MOST grant NSC 101-2112-M-008-017-MY3 and 104-2112-M-008-012-MY3.  W.~P.~C. is supported by the MOST grants 103-2112-M-008-024-MY3. 

\bibliography{../reference}

\end{document}